\documentclass[a4paper, aps, superscriptaddress, floatfix, twocolumn, longbibliography]{revtex4-1}

\usepackage{amsmath,amssymb}
\usepackage{graphicx}
\usepackage{dcolumn}
\usepackage{bm}
\usepackage[utf8]{inputenc}
\usepackage[caption=false]{subfig}
\usepackage{floatrow}
\usepackage{float}
\usepackage[rgb]{xcolor}
\usepackage{epsfig}

\newcommand{\SM}[1]{SM}

\begin{document}

\title{$k$-core structure of real multiplex networks}

\author{Saeed Osat}\email{saeedosat13@gmail.com}
\affiliation{Deep Quantum Labs, Skolkovo Institute of Science and Technology, Moscow 143026, Russia}

\author{Filippo Radicchi}
\affiliation{Center for Complex Networks and Systems Research, Luddy School
  of Informatics, Computing, and Engineering, Indiana University, Bloomington,
  Indiana 47408, USA}

\author{Fragkiskos Papadopoulos}
\affiliation{Department of Electrical Engineering, Computer Engineering and Informatics, Cyprus University of Technology, 33 Saripolou Street, 3036 Limassol, Cyprus}

\date{\today}

\begin{abstract}
Multiplex networks are convenient mathematical representations for many real-world -- biological, social, and technological -- systems of interacting elements, where pairwise interactions among elements have different flavors. Previous studies pointed out that real-world multiplex networks display significant inter-layer correlations -- degree-degree correlation, edge overlap, node similarities --  able to make them robust against random and targeted failures of their individual components. Here, we show that inter-layer correlations are important also in the characterization of their $\mathbf{k}$-core structure, namely the organization in shells of nodes with increasingly high degree. Understanding $k$-core structures is important in the study of spreading processes taking place on networks, as for example in the identification of influential spreaders and the emergence of localization phenomena.  We find that, if the degree distribution of the network is heterogeneous, then a strong $\mathbf{k}$-core structure is well predicted by significantly positive degree-degree correlations. However, if the network degree distribution is homogeneous, then strong $\mathbf{k}$-core structure is due to positive correlations at the level of node similarities. We reach our conclusions by analyzing different real-world multiplex networks, introducing novel techniques for controlling inter-layer correlations of networks without changing their structure, and taking advantage of synthetic network models with tunable levels of inter-layer correlations.
\end{abstract}
                           
\maketitle

\section*{Introduction}
 
A multiplex network is a collection of single-layer networks sharing common nodes, where each layer captures a different type of pairwise interaction among nodes~\cite{GinestraBook,BOCCALETTI20141, Lee2015, Kivela2014,MultNetMathPRX}. This is a convenient and meaningful representation for many real-world networked systems, including social~\cite{Szell2010,Mucha2010}, technological~\cite{ark2009}, and biological systems~\cite{Bullmore2009, manlio2015, lim2019discordant}. The simultaneous presence of different types of interactions is at the root of the observation of collective phenomena generally not possible in single-layer networks. A paradigmatic example is provided in the seminal study by Buldryev \emph{et al}.~\cite{Buldyrev2010} where it was shown that, if multiplexity is interpreted as a one-to-one interdependence among corresponding nodes in the various layers, then the mutual connectedness of  a multiplex network displays an abrupt breakdown under random failures of its nodes. Other examples of anomalous behavior of multiplex networks regard both dynamical and structural processes~\cite{spreadingreview2016, Baxter2012PRL, Radicchi2013NatPhys, Radicchi2015NatPhys, RadicchiBianconi2017PRX, Osat2017NatComm, Osat2018PhysA, BaxterTarget2018PRE}.  Although multiplexity seems a necessary condition for the emergence of non-trivial collective behavior, the magnitude of the anomalous behavior in real-world multiplex networks is often suppressed by the presence of strong inter-layer correlations, such as link overlap, degree-degree correlations, geometric correlations and correlated community structure~\cite{Reis2014NatPhys, Nicosia2015PRE, Radicchi2015NatPhys, KleinebergNatPhys, KleinebergPRL, FaqeehPRL}.

An important feature characterizing structural and dynamical properties of single-layer networks is the so-called $k$-core structure~\cite{kcore,Gkcore}. The $k$-core of a network is the maximal subgraph of the network in which all vertices have degree at least $k$ (see Methods section~\ref{methods_core}). The notion of $k$-core is used to define so-called $k$-shells of nodes, and further to define a node centrality metric $k_\textnormal{s}$  named $k$-shell index or coreness (Methods section~\ref{methods_core}). $k$-cores, and $k$-shells, are particularly important for the understanding of spreading processes on networks~\cite{pastor2015epidemic}. For instance,  the coreness of a node is a good indicator of its spreading power~\cite{Kitsak2010}. Also, in many real-world networks, the notion of maximal $k$-core, i.e., the core with the largest $k$, represents a good structural proxy for the understanding of dynamical localization phenomena in spreading processes~\cite{pastor2018eigenvector}. Finally,
the extinction of species located in the maximal $k$-core well predicts the collapse of networks describing mutualistic ecosystems~\cite{Morone2019}.

The notion of $k$-core can be generalized to the case of multiplex networks~\cite{AzimiKcore2014PRE}. In a multiplex of $L$~layers, the $\mathbf{k}$-core is defined for a vector of degree threshold values $\mathbf{k}  = (k_1, \ldots ,  k_\ell , \ldots, k_L)$. Specifically, it is the maximal set of nodes such that each node complies with the corresponding degree threshold condition in each layer of the multiplex (Methods section~\ref{methods_core}). In Ref.~\cite{AzimiKcore2014PRE}, Azimi-Tafreshi and collaborators studied the emergence of $\mathbf{k}$-cores in random uncorrelated multiplex network models with arbitrary degree distributions. They showed that $\mathbf{k}$-cores in multiplex networks are characterized by abrupt transitions, but their properties cannot be easily deduced from those of the $k$-cores of the individual network layers. They further studied the $\mathbf{k}$-core structure of a few real-world networks. They noted that these systems display significant differences from the theoretical predictions that can be obtained in the framework developed for uncorrelated networks, thus indicating the necessity of a better understanding of the role of structural correlations in the characterization of the  $\mathbf{k}$-core structure of real-world multiplex networks.

In this paper, we build on the work of  Azimi-Tafreshi {\it et al.}~\cite{AzimiKcore2014PRE} and perform a systematic characterization of the $\mathbf{k}$-core structure of real-world multiplex networks. We consider a large variety of systems, and study how the size of the $\mathbf{k}$-core depends on the choice of the vector $\mathbf{k}$. Specifically, we compare  the $\mathbf{k}$-core of real-world networks with the core observed for the same choice of the vector $\mathbf{k}$ on randomized versions of the networks where inter-layer correlations are destroyed. We find that real-world multiplex networks possess non-null $\mathbf{k}$-cores while their reshuffled versions do not. We interpret this fact as a sign of the strength of the $\mathbf{k}$-core structure of real-world multiplex networks. To provide an intuitive explanation of this finding, we take advantage of the geometric interpretation of inter-layer correlations in terms of network hyperbolic embedding~\cite{Krioukov2010PRE, hypermap_cn}. Our choice is motivated by a series of recent studies where it has been shown that not only real-world multiplex networks display significant geometric correlations~\cite{KleinebergNatPhys}, but also that the amount of these correlations is a good predictor of the robustness of the system under targeted attacks~\cite{KleinebergPRL, FaqeehPRL}.
In network hyperbolic embedding, nodes of a network are mapped to points of the two-dimensional hyperbolic disk~\cite{Papadopoulos2012Nature}. The radial coordinate of a node in the disk quantifies the popularity of the node; the difference between angular coordinates is related instead to the level of similarity between pairs of nodes. Geometric correlations in a multiplex network are quantified by looking at the coordinates of the same node in different layers, provided that the layers are embedded independently in the hyperbolic space. Geometric correlations can be quantified either for radial or angular coordinates of the nodes. Both types of correlations are able to provide insights about the $\mathbf{k}$-core structure of a multiplex. Specifically, we show that the more heterogeneous are the degree distributions of the layers, the more pivotal is the role of popularity correlations in the emergence of strong $\mathbf{k}$-core structure. On the other hand, the less heterogeneous are the degree distributions, the more crucial is the role of similarity correlations. These observations are in remarkable agreement with the behavior observed in synthetic multiplex networks where we can control the level of geometric correlations
across the layers~\cite{KleinebergNatPhys}.

\section*{Results}

\subsection*{Single-layer networks}

We start by studying the $k$-core structure of single-layer networks. Most of our results for single-layer networks are not novel as the problem
was already studied in Ref.~\cite{Boguna2008}.  We replicate and expand the analysis of Ref.~\cite{Boguna2008} here for two main reasons.
First, the repetition of the analysis of Ref.~\cite{Boguna2008} allows us to have a self-contained paper. Second and more important,  the analysis serves to properly calibrate our framework before extending it to the study of the $\mathbf{k}$-core structure of multiplex networks. Such a calibration is of fundamental importance as findings on single-layer networks provide us with proper baselines for the interpretation of results valid for multiplex $\mathbf{k}$-core structures, including testable hypotheses on their expected behavior.

\begin{figure}[]
\captionsetup{farskip=0pt}
\includegraphics[width=1\columnwidth]{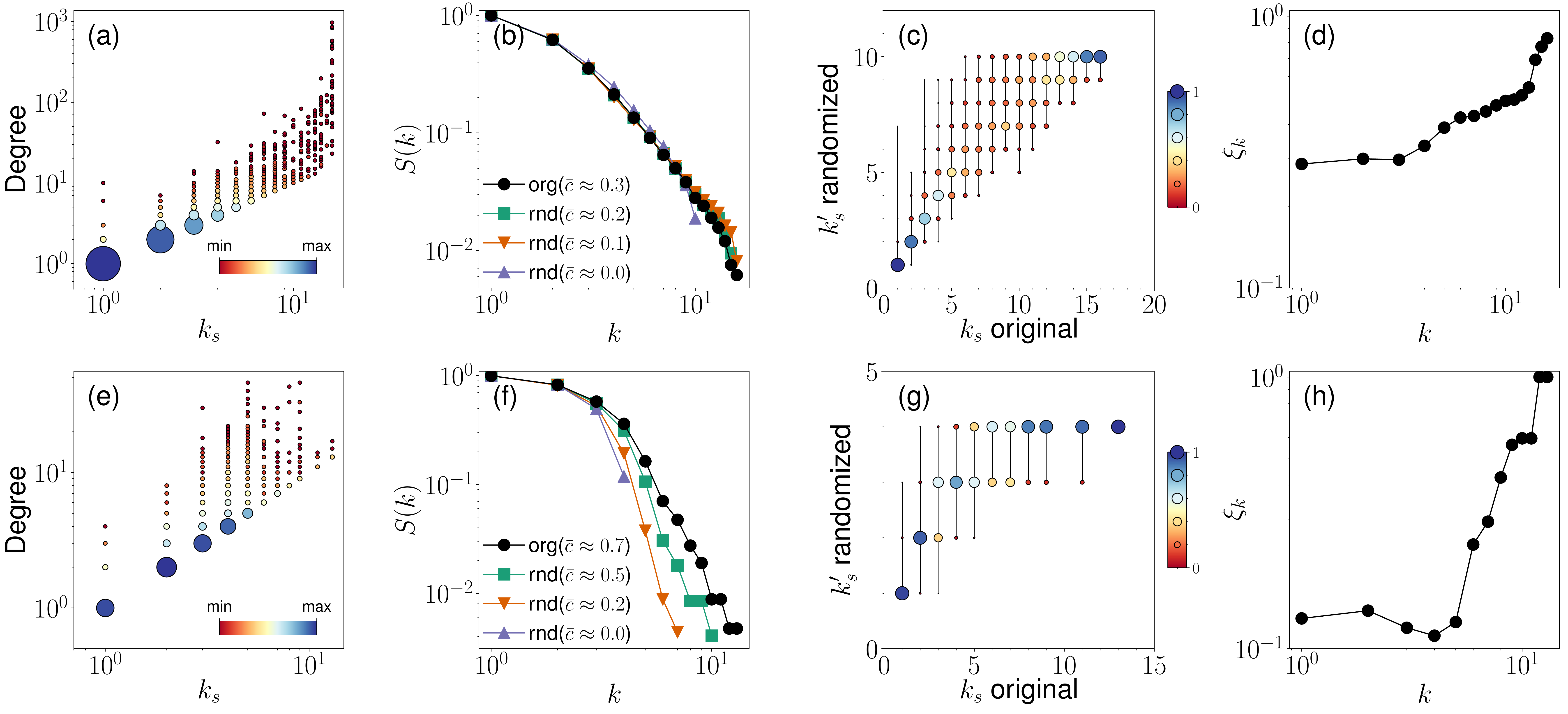}
 \caption{ {\bf $k$-core structure of real-world networks.} We analyze single-layer networks. The top row refers to results valid for the IPv6 Internet; the bottom row refers to results valid for the arXiv co-authorship network. ({\bf a} and {\bf e})  Scatter plot of node degrees vs. coreness. The size of the symbols is proportional to the number of nodes having each specific degree and $k$-shell index values. ({\bf b} and {\bf f}) Relative size $S(k)$ of the $k$-core (see Methods section~\ref{methods_core}) in the real networks (labeled as ``org'') and their randomized counterparts (labeled as ``rnd''). Randomized networks are obtained by shuffling random pairs of edges while controlling for the average value of the clustering coefficient $\bar{c}$ (Methods section~\ref{methods_rand}). $\bar{c} \approx 0$ is obtained after $10,000$ and $2,000$ link rewirings in the Internet and arXiv, respectively. ({\bf c} and {\bf g}) $k$-shell index of nodes before and after network randomization (obtained for $\bar{c}\approx 0$). The size of the symbols is proportional to the percentage of nodes whose coreness changed from $k_s$ in the original network to $k_s'$ in the reshuffled network.  ({\bf d} and {\bf h})  Angular coherence $\xi_k$ of the nodes belonging to each $k$-core.}
\label{fig1}
\end{figure}

In Figure~\ref{fig1}, we report results obtained by analyzing two single-layer networks: a snapshot of the Internet at the IPv6 level~\cite{INTERNET} and the co-authorship network formed by the authors of papers in the ``Biological Physics" category of  arXiv~\cite{Manlio2015PRX}. Details on the data and results for other networks can be found in Supplemental Material~\cite{SM}, sections~I and~II. The $k$-shell index of the nodes is strongly correlated with their degree (Figures~\ref{fig1}a and~\ref{fig1}e and Supplemental Material~\cite{SM}, Figure~2a). However, as previously noted in Ref.~\cite{Kitsak2010}, nodes with the same value of the $k$-shell index may correspond to very different degree values. Further, we note that the degree distribution of the Internet is much broader than the one of the arXiv (see  Figures~\ref{fig1}a~and~\ref{fig1}e and Supplemental Material~\cite{SM}, Figure~1). Specifically, the degree distributions of both networks can be modeled quite well in terms of power laws, i.e., $P(k) \sim k^{-\gamma}$, with degree exponent $\gamma = 2.1$ for the Internet and $\gamma = 2.6$ for the arXiv, thus indicating that the degree distribution of the Internet is more heterogeneous than the one of the arXiv. The correlation between $k$-shell index and node degree weakens significantly as we move into inner $k$-shells in the arXiv, but not in the Internet (Supplemental Material~\cite{SM}, Figure~2a). We have verified that the less heterogeneous is the degree distribution the weaker is the correlation between
$k$-shell index and degree (Supplemental Material~\cite{SM}, Figure~2b).

To quantify the quality of the $k$-core structure we consider the relative size $S(k)$ of the $k$-core as a function of the value of the threshold $k$.  If there is a rich collection of $k$-cores with a wide spectrum of $k$'s, then the $k$-core structure is strong; it is weak, otherwise. Figures~\ref{fig1}b~and~\ref{fig1}f show that the $k$-core structures of the Internet and arXiv are strong. In particular, we see that $S(k)$ decreases smoothly as $k$ increases, while $S(k) > 0$ up to $k=16$ for the Internet, and up to $k=13$ for the arXiv. 

Ref.~\cite{Boguna2008} showed in experiments with synthetic networks that both degree heterogeneity and clustering improve the quality of the $k$-core structure. To study how these properties affect the quality of the $k$-core structure of real networks, we study the behavior of $S(k)$ on degree-preserving randomized versions of the networks. The randomization is performed by rewiring randomly chosen links till the value of the average clustering in the network is reduced to a pre-defined value (see Methods section~\ref{methods_rand}).  We see in Figures~\ref{fig1}b~and~\ref{fig1}f that the randomization affects the $k$-core structure of the Internet to a much lesser extent than the $k$-core structure of the arXiv, while the effect is stronger the more we destroy clustering. As Figures~\ref{fig1}c~and~\ref{fig1}g clearly show, the effect of the randomization consists in redistributing nodes to lower $k$-shell values. Specifically, these figures show the percentage of nodes, indicated by the circles, whose $k$-shell index changes from $k_s$ in the original network ($x$-axis) to $k_s'$ in the randomized network ($y$-axis). We see that changes of the $k$-shell values induced by the randomization are much more apparent for the  arXiv than in the Internet---nodes in the arXiv are redistributed to significantly lower shells. For instance, we see in Figure~\ref{fig1}g that nodes belonging to $k_s=11$ in the original network move to $k_s'=4$ and $k_s'=3$ in the randomized network. These results indicate that networks with more heterogeneous degree distributions can have strong $k$-core structures even if their clustering is weak. On the other hand, if the degree distribution is less heterogeneous, clustering becomes more important for having a strong $k$-core structure. In the next section we explicitly verify these observations in controlled experiments with synthetic networks (Figures~\ref{fig2}e-g).

\subsection*{Hyperbolic embedding}

To better capture the role of correlations for the characterization of the $k$-core structure of networks, we decided to take advantage of the vectorial representation of nodes in the hyperbolic space~\cite{Krioukov2010PRE, Boguna2010, Papadopoulos2012Nature}. According to this mapping, every node $i$
of a network becomes a point, identified by the coordinates $(r_i, \theta_i)$, in the two-dimensional hyperbolic disk (see Methods sections~\ref{methods_S1} and~\ref{methods_embedding}). The radial coordinate $r_i$ quantifies the popularity of node $i$ in the network, and basically corresponds to the degree $k_i$ of the node (Methods section~\ref{methods_embedding}).  The angular coordinate $\theta_i$ serves to quantify pairwise similarities, in the sense that the angular distance between pairs of nodes is inversely proportional to their similarity.  Whereas radial coordinates don't convey more explicative information than node degrees,  angular coordinates offer the opportunity to deal with node similarities in continuous space, thus allowing for smooth and easily quantifiable metrics of similarities of arbitrary sets of nodes, including $k$-cores. Specifically, we use a measure of coherence among angular coordinates of nodes within the $k$-core, namely $\xi_k$, to measure the average level of similarity among the nodes within the $k$-core~\cite{FaqeehPRL} (see Methods section~\ref{methods_coherence}).
By definition $\xi_k \in [0,1]$, with $\xi_k =0$ meaning that the angular coordinates of the $k$-core are uniformly scattered around the disk, and $\xi_k = 1$ meaning that all nodes within the $k$-core have identical value for their angular coordinates.
Figures~\ref{fig1}d and~\ref{fig1}h show $\xi_k$ as a function of $k$ for the Internet and arXiv networks, respectively. We see that $\xi_k$ increases with $k$, meaning that as we move to inner $k$-cores, angular coordinates of the nodes tend to be more localized. Similar results hold if one analyzes other real networks and if one measures angular coherence in the $k$-shells instead of the $k$-cores (see Supplemental Material~\cite{SM}, section~II).

We take advantage of network hyperbolic embedding not only for descriptive purposes, but also to perform controlled experiments. We leverage  models introduced in the literature on network hyperbolic embedding to  better understand the role played by clustering and node similarities in predicting the strength of network $k$-core structure. Specifically, we rely on network instances generated according to the $\mathbb{S}^{1}$ model~\cite{Krioukov2010PRE, Serrano2008}, which is isomorphic to hyperbolic geometric graphs (see Methods section~\ref{methods_S1}). The model generates synthetic networks with arbitrary degree distribution and clustering strength. 

\begin{figure}[]
\captionsetup{farskip=0pt}
\includegraphics[width=0.8\columnwidth]{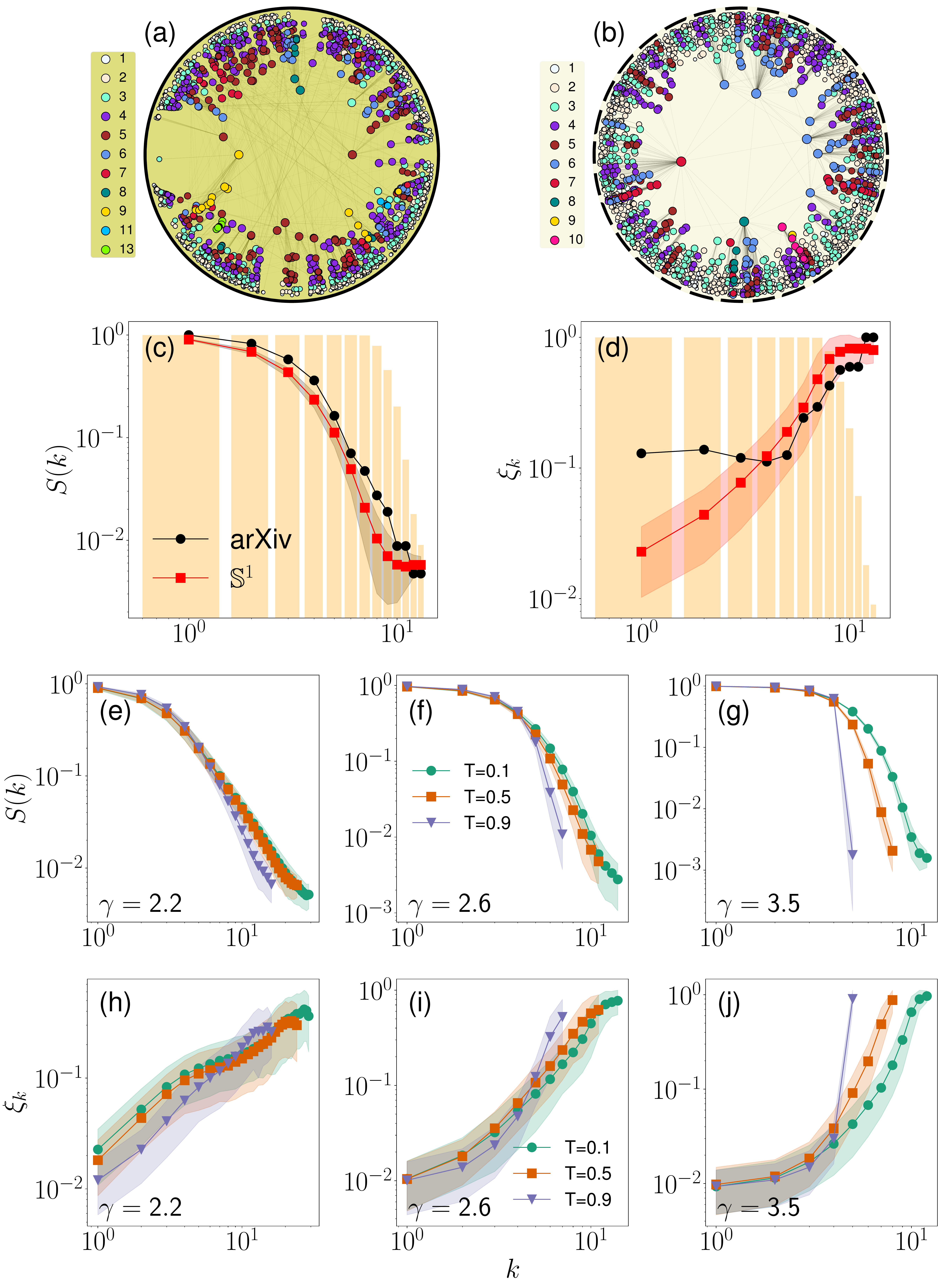}
\caption{ {\bf $k$-core structure of single-layer networks.} ({\bf a}) Hyperbolic embedding of the arXiv network. The position of the nodes in the disk is determined by their hyperbolic coordinates; different colors serve to differentiate nodes depending on their $k$-shell index value. ({\bf b}) Same as in panel a, but for an instance of the $\mathbb{S}^{1}$ model built using similar characteristics as in the  arXiv network (i.e., same network size $N$, and same values of the degree exponent $\gamma$,  average degree $\bar{k}$, and average clustering coefficient $\bar{c}$).
 ({\bf c}) Relative size $S(k)$  of the $k$-core as a function of the threshold value $k$ for the arXiv netwok and the $\mathbb{S}^{1}$ model. 
 The results for the modeled network are average values over $1,000$ network instances.
 The shaded area identifies the region corresponding to one standard deviation away from the average. The average value is computed only over non-null $k$-cores, and the bars in the background of the figure display the fraction of instances where such non-empty cores were indeed present.  ({\bf d}) We consider the same data as in panel c, but monitor the angular coherence $\xi_k$ as a function of $k$. ({\bf e}) $S(k)$ vs. $k$ for the $\mathbb{S}^{1}$. We set here the size of the network $N=10,000$, degree exponent $\gamma = 2.2$, and average degree $\bar{k}=6$. We consider three different values of the  temperature parameter $T$. This serves to tune the average value  of the clustering coefficient $\bar{c}$ of the model, as $T$ is inversely proportional to $\bar{c}$. Results  are averaged over $200$ instances of the model. Shaded areas stand for one standard deviation away from the average. ({\bf f}) Same as in panel e, but for $\gamma = 2.6$. ({\bf g}) Same as in panel e, but for $\gamma = 3.5$.
 ({\bf h}) We consider the same networks as in panel e, but we monitor angular coherence $\xi_k$ vs. $k$. ({\bf i}) Same as in panel h, but for $\gamma = 2.6$. ({\bf j}) Same as in panel h, but for $\gamma = 3.5$.}
 \label{fig2}
\end{figure}

In  Figure~\ref{fig2}, we perform a direct comparison between the relative size $S(k)$ and angular coherence $\xi_k$ of the $k$-core structure of the arXiv collaboration network and of a synthetic graph generated according to the $\mathbb{S}^{1}$ model with similar  values of number of nodes, average degree, and average clustering coefficient as of the arXiv collaboration network. The synthetic network has a power-law degree distribution $P(k)\sim k^{-\gamma}$ with exponent $\gamma=2.6$, compatible with the one of the real-world network (Supplemental Material~\cite{SM}, section~I). We see that the two graphs display a qualitatively similar behavior with respect to $S(k)$ (Figure~\ref{fig2}c) and $\xi_k$ (Figure~\ref{fig2}d) as functions of the threshold value $k$.

Synthetic networks allow us to play with the ingredients that we believe are important in the characterization of network $k$-core structure. We see that the range of $k$ values for which we have non-null $k$-cores widen not only when the degree distribution becomes more heterogeneous (lower $\gamma$ values), but also  when the clustering coefficient increases (Figures~\ref{fig2}e-g). In all these cases, nodes belonging to inner $k$-cores always have more similar angular coordinates in the hyperbolic embedding (Figures~\ref{fig2}h-j).

\subsection*{Multiplex networks}

We now turn our attention to the study of the $\mathbf{k}$-core structure of real-world multiplex networks. For simplicity, we limit our attention to two-layer multiplex networks only, so that $\mathbf{k} = (k_1, k_2)$.  We note that a necessary condition for having a non-null $(k_1, k_2)$-core is that the $k_1$-core of layer $\ell =1$ and the  $k_2$-core of layer $\ell =2$ are simultaneously non null. 
The condition is clearly not sufficient, as there could be combinations $(k_1, k_2)$ associated to empty cores in the multiplex but still showing non-empty cores at the level of the individual layers. As a consequence, we expect that multiplex networks displaying low inter-layer correlation at the node level will be weak in terms of  $\mathbf{k}$-core structure, in the sense that non-empty cores will exist only for limited choices of the thresholds $(k_1, k_2)$. Based on our knowledge of the relation between $k$-core strength and hyperbolic network embedding, we further expect that inter-layer correlations that are important in the prediction of the strength of the $\mathbf{k}$-core structure of a multiplex are not only those relative to the degree of the nodes, but also those concerning the similarity among pairs of nodes.

\begin{figure}[]
\captionsetup{farskip=0pt}
\includegraphics[width=1\columnwidth]{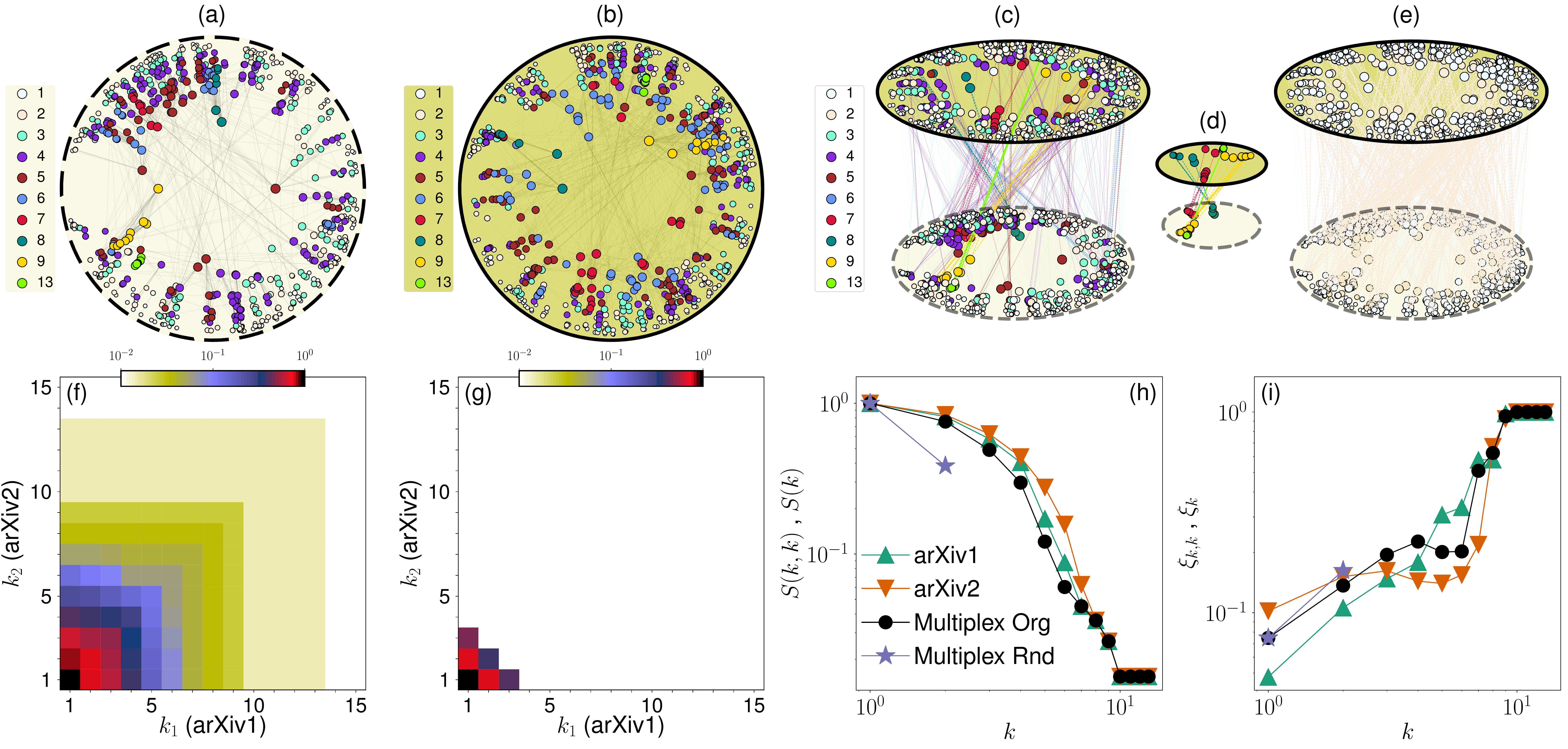}
\caption{ {\bf $\mathbf{k}$-core of real-world multiplex networks.}
({\bf a} and {\bf b}) Hyperbolic embedding of the arXiv multiplex network. Panel a refers to the layer arXiv1, while panel b to  the layer arXiv2. The position of the nodes in the disk is determined by their hyperbolic coordinates, and only nodes that exist in both layers are shown ($911$ nodes); different colors serve to differentiate nodes depending on their $k$-shell index value.
({\bf c}) Correspondence among nodes belonging to the $(k, k)$-shells (see Methods section~\ref{methods_core}) of the arXiv multiplex network. ({\bf d}) Same as in panel c, but for $k \geq 7$. ({\bf e}) Same as in panel c, but for the randomized version of the multiplex where the node labels of one of the two layers are randomly reshuffled.
({\bf f}) Relative size $S(k_1, k_2)$ of the $(k_1, k_2)$-core for the arXiv multiplex network. ({\bf g}) Same as in panel f, but for the randomized version of the multiplex network. ({\bf h}) Relative size $S(k, k)$ of the $(k, k)$-core for the arXiv multiplex network, and its randomized version. These curves are compared with those of the relative size $S(k)$ of the $k$-core of the individual layers. ({\bf i}) Same as in panel h, but for the metrics of angular coherence $\xi_{k, k}$ and $\xi_{k}$.}
\label{fig3}
\end{figure}

In Figure~\ref{fig3}, we consider a multiplex version of the  arXiv collaboration network, where one layer is obtained by considering manuscripts of the section ``Biological Physics'' (i.e., the one considered already in Figures~\ref{fig1} and~\ref{fig2}), and the other based on manuscripts of the section ``Data Analysis, Statistics and Probability."  For sake of brevity, we will refer to them as arXiv1 and arXiv2, respectively.
We observe that  the $\mathbf{k}$-core structure of the multiplex network is quite robust, in the sense that the relative size $S(k_1, k_2) $ of the $(k_1, k_2)$-core is strictly larger than zero for a wide range of choices of the threshold values $(k_1, k_2)$ (Figure~\ref{fig3}f). This fact becomes apparent when the results valid for the real network are contrasted with those valid for a randomized version of the network (Figure~\ref{fig3}g). The randomization here consists of randomly shuffling the labels of the nodes of one of the two layers, so that the topology of both layers remains unchanged, but inter-layer correlations are completely destroyed (Methods section~B).  As a visual inspection of Figures~\ref{fig3}f and \ref{fig3}g reveals, the real network displays non-empty cores  in a much wider  region of the $(k_1, k_2)$ plane than the randomized version of the network.  The result is highlighted in Figure~\ref{fig3}h for the special case $k_1 = k_2 =k$,  where we see that the $S(k, k)$ of the real-multiplex network behaves almost identically to the $S(k)$ of the individual layers. On the contrary, the randomized version of the multiplex network displays an empty core already for $k>2$. We can interpret the robustness of the $\mathbf{k}$-core of the real multiplex network in terms of inter-layer correlations. Indeed in Figure~\ref{fig3}i, we see that nodes belonging to inner cores have simultaneously high angular coherence $\xi_{k, k}$ (Methods section~\ref{methods_coherence}) in both layers of the real multiplex, a situation visualized in Figures~\ref{fig3}c and \ref{fig3}d vs. Figure~\ref{fig3}e for the randomized version of the network. Similar results hold for other real-world multiplex networks (Supplemental Material~\cite{SM}, section~III).  


Next, we investigate the extent to which degree and similarity correlations affect the $\mathbf{k}$-core structure,  separately. To this end, we take advantage of network hyperbolic embedding, where layers are embedded independently, thus each node has radial and angular coordinates for each layer of the multiplex. Also in this case, we consider the degree of the nodes instead of their radial coordinate, being the two quantities clearly related one to  the other. We break each type of correlation while preserving the other type of correlation. To break degree correlations we consider the common nodes in the two layers of the multiplex, i.e., the nodes that are simultaneously present in both layers. Then, we select one of the layers and sort the common nodes with respect to their angular coordinates. We group the nodes in consecutive groups of size $n$, and in each group we reshuffle node labels. If $n$ is sufficiently small, correlations among angular coordinates are approximately preserved since the angular coordinates of nodes do not change significantly within the group. Clearly, for $n=1$ no reshuffling is performed, while if $n=N$, where $N$ is the number of common nodes, then all types of inter-layer correlations are broken. To break correlations among angular coordinates while preserving degree correlations we follow a similar procedure. Specifically, we select one of the layers, sort the common nodes with respect to their degrees, group nodes in consecutive groups of size $n$, and reshuffle node labels in each group.

\begin{figure}[]
\captionsetup{farskip=0pt}
\includegraphics[width=1\columnwidth]{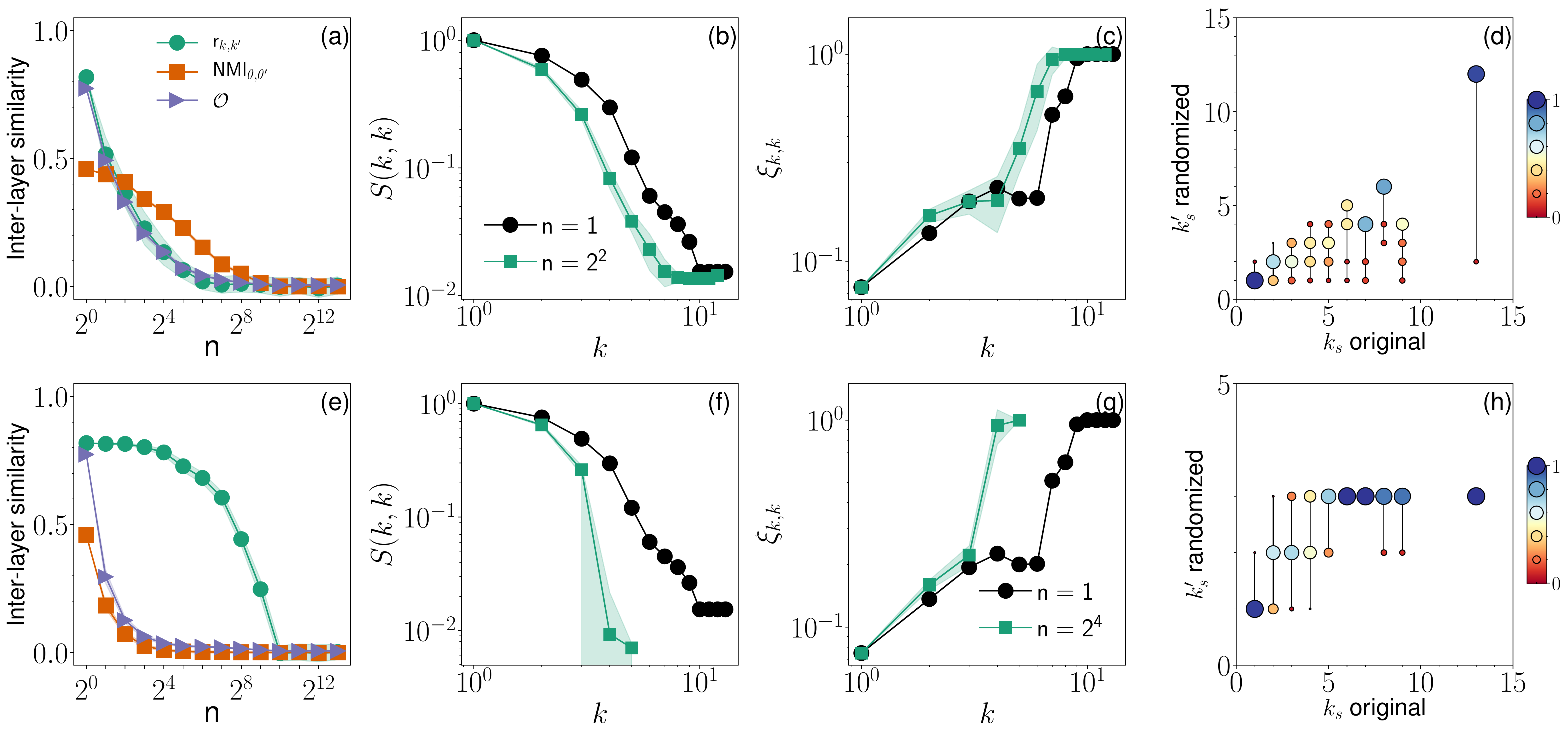}
\caption{ {\bf Inter-layer correlations and the $\mathbf{k}$-core structure of the arXiv multiplex network.} We analyze the arXiv multiplex network.
  ({\bf a}) Different metrics of inter-layer similarity as a function of the group size $n$ used to randomize node labels, thus breaking inter-layer degree correlations.  For $n=1$, node labels of the network are not randomized; full shuffle of node labels is obtained for large $n$ values.   We focus here on the case where degree inter-layer correlation is broken, but we preserve inter-layer correlation among nodes' angular coordinates (see main text for details).
  Metrics of similarities considered here are the Pearson correlation coefficient $r_{k,k'}$ among the degrees of nodes in the two layers;  normalized mutual information $\textnormal{NMI}_{\theta, \theta'}$ of the angular coordinates of the nodes in the two layers; and edge overlap $\cal{O}$ among the two layers (Methods section~\ref{methods_similarity}).
  ({\bf b}) Relative size $S(k,k)$ of the $(k,k)$-core. The results of the original multiplex network ($n=1$) are compared with those valid for $n=4$. At this level of randomization, we find that $r_{k,k'}=0.36$ and $\textnormal{NMI}_{\theta, \theta'}=0.41$. These numbers should be compared respectively with  $r_{k,k'}=0.82$ and
  $\textnormal{NMI}_{\theta, \theta'}=0.46$ of the original network. The results for $n=4$ are average values obtained on $100$ independent randomizations. Shaded areas identify the region corresponding to one standard deviation away from the average.   ({\bf c}) Same as in panel b, but for the angular coherence $\xi_{k,k}$. ({\bf d}) Scatter plot of the $(k,k)$-shell index of nodes in the original vs. the randomized multiplex network. The size of the symbols is proportional to the percentage of points in the scatter plot. ({\bf e}, {\bf f}, {\bf g} and {\bf h}) Same as in panel a, b, c and d, respectively. We consider here the case where inter-layer correlation among nodes' angular coordinates is destroyed, but inter-layer correlation among nodes' degrees is preserved (see main text for details). The results of the original network are compared with those obtained for $n=16$, when $r_{k,k'}=0.78$ and $\textnormal{NMI}_{\theta, \theta'}=0.01$.}
\label{fig4}
\end{figure}

\begin{figure}[]
\captionsetup{farskip=0pt}
\includegraphics[width=1\columnwidth]{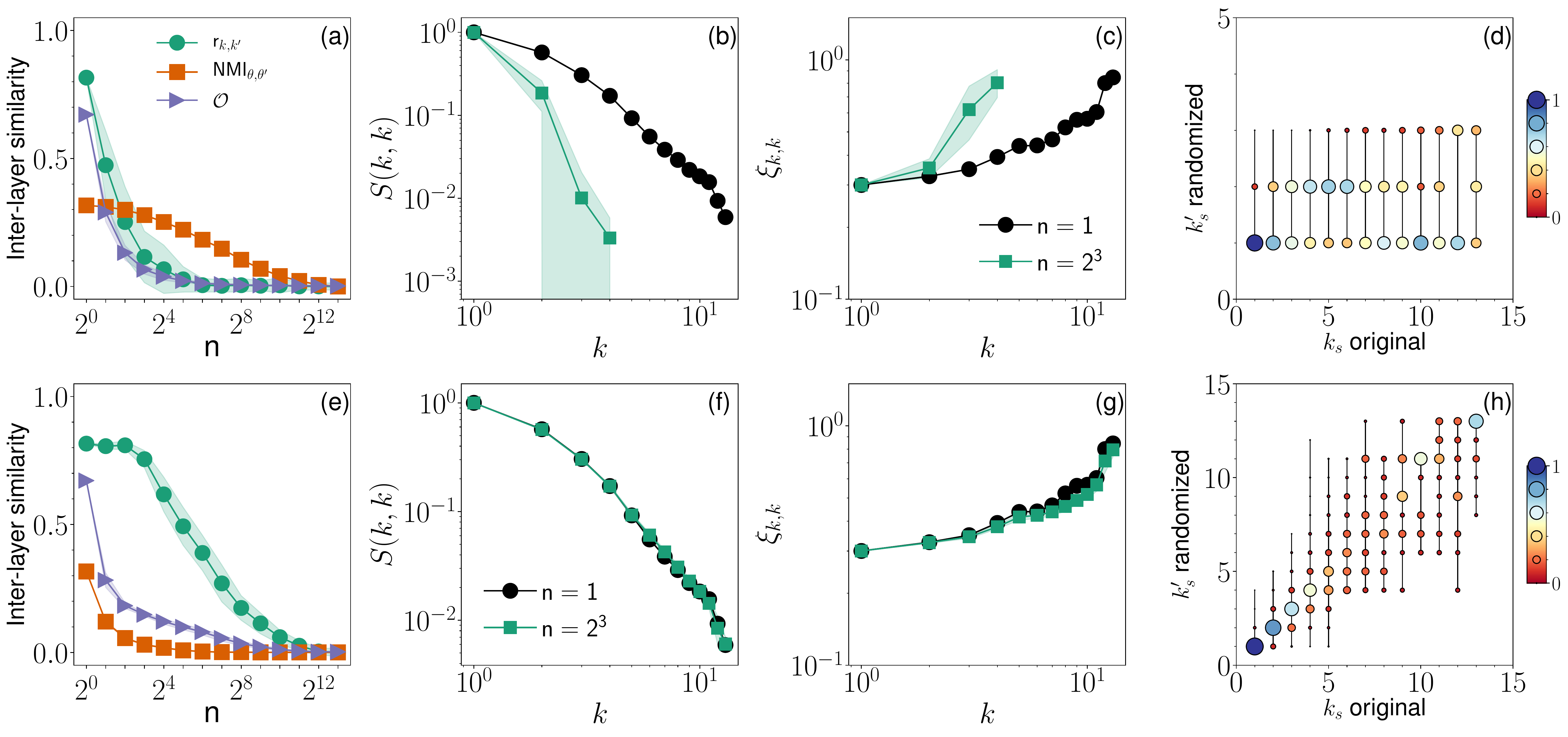}
\caption{{\bf Inter-layer correlations and the $\mathbf{k}$-core structure of the Internet multiplex network.}
 Same analysis as in Figure~\ref{fig4}, but for the IPv4/IPv6 Internet multiplex network.
Correlations of the original network are such that $r_{k,k'}=0.82$ and $\textnormal{NMI}_{\theta, \theta'}=0.32$. Results of the real-world system are compared with those obtained after destroying inter-layer degree correlations such that $r_{k,k'}=0.12$ and $\textnormal{NMI}_{\theta, \theta'}=0.28$  in the top-row panels, and after destroying angular correlations such that $r_{k,k'}=0.76$ and $\textnormal{NMI}_{\theta, \theta'}=0.03$ in the bottom-row panels.}
\label{fig5}
\end{figure}

The top row of Figure~\ref{fig4} shows the results valid for the arXiv multiplex network when degree correlations are broken while correlations among angular coordinates are preserved; the bottom row of Figure~\ref{fig4} reports results valid when degree correlations are preserved, but correlations among angular coordinates are destroyed.  As expected, inter-layer degree correlation, measured in terms of Pearson correlation coefficient $r_{k, k'}$ (see Methods section~\ref{methods_similarity}), decreases with the size $n$ of the groups used in the randomization procedure (Figure~\ref{fig4}a).  Similarly, correlation among angular coordinates of the nodes, measured in terms of the normalized mutual information $\textnormal{NMI}_{\theta, \theta'}$ (Methods section~\ref{methods_similarity}), decreases as $n$ increases. There is, however, a range of $n$ values where $r_{k, k'}$ is low and $\textnormal{NMI}_{\theta, \theta'}$ high, indicating that correlation at the level of angular coordinates is preserved but degree correlation is destroyed. We consider the randomized version of the network obtained for $n=4$, thus belonging to the aforementioned range of suitable $n$ values, and study differences between its $(k, k)$-core structure and the one of the real multiplex network (Figures~\ref{fig4}b and~\ref{fig4}c). The $(k, k)$-core of the real network is only slightly more robust than the one of the randomized network (Figure~\ref{fig4}b). Angular coordinates of the nodes in the inner cores are still strongly correlated (Figure~\ref{fig4}c). The same analysis gives a completely different result in the case of the Internet multiplex network, where the two layers are given by the IPv4 and IPv6 topologies, respectively (see Supplemental Material~\cite{SM}, section~I for details on the data). Reducing degree correlation in this case, destroys the $(k, k)$-core structure (Figure~\ref{fig5}b-d).

If we repeat the same exercise, but now destroying correlations among angular coordinates while preserving correlations between degrees, we see a completely different picture. For the arXiv multiplex network, the randomization procedure leads to the destruction of the $\mathbf{k}$-core structure (Figures~\ref{fig4}f-h). Instead, for the Internet multiplex network, we see that the randomization procedure has virtually no effect on the strength of the $\mathbf{k}$-core structure, keeping it unchanged with respect to the one of the original network (Figures~\ref{fig5}f-h).

\begin{figure}[]
\captionsetup{farskip=0pt}
\includegraphics[width=0.45\columnwidth]{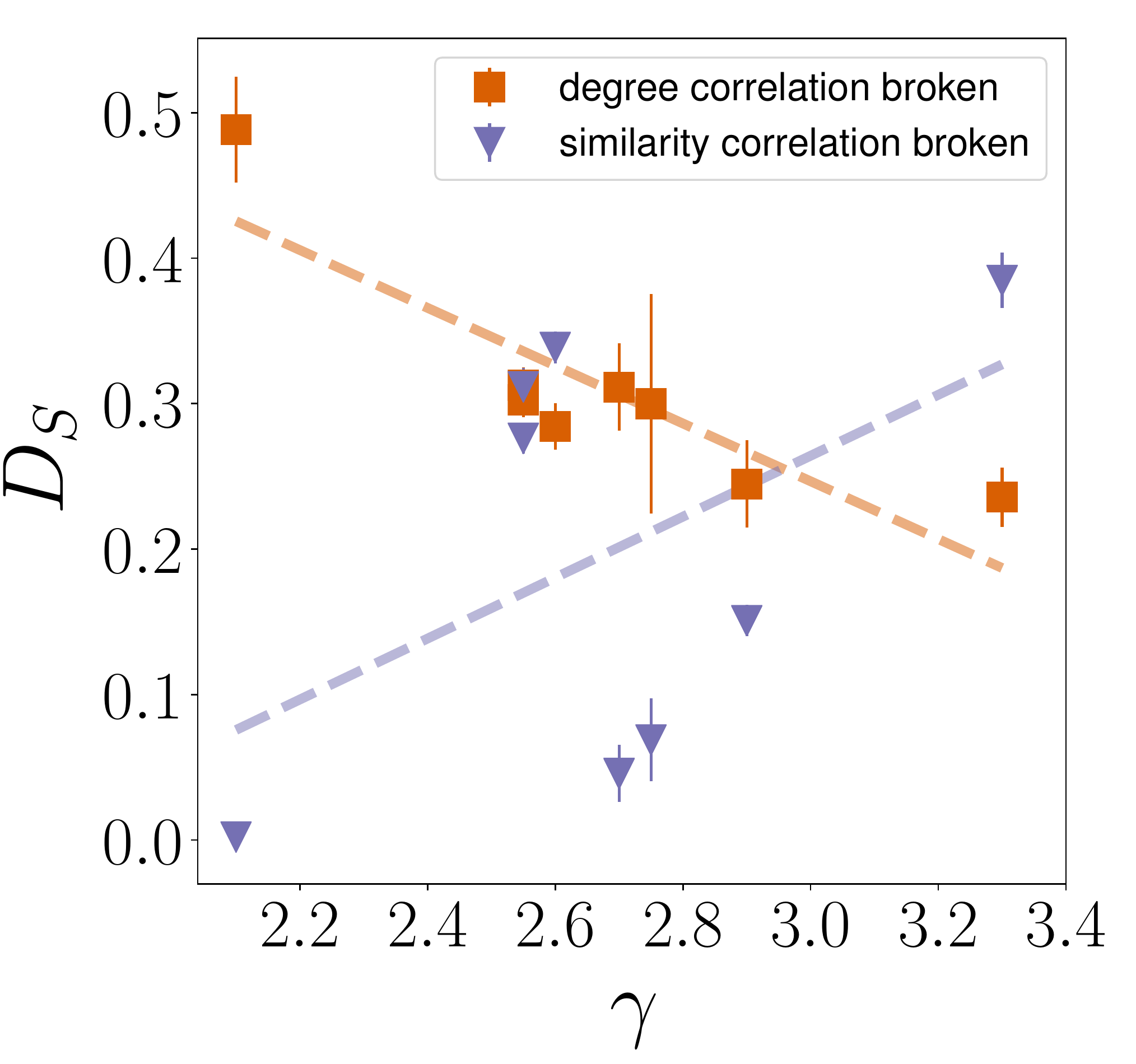}
\caption{{\bf Quantifying the effect of inter-layer degree and similarity correlations in the $\mathbf{k}$-core structure of real-world multiplex networks.} Relative difference $D_S = \left[\sum_k S(k, k)^\textnormal{org}-\sum_k S(k, k)^\textnormal{rnd}\right]/\sum_k S(k,k)^\textnormal{org}$ between the relative $(k, k)$-core sizes $S(k, k)^\textnormal{org}$ and $S(k, k)^\textnormal{rnd}$ of real-world multiplex networks and their randomized counterparts. In the randomized counterparts, we either destroy inter-layer degree correlation or correlation among the nodes' angular coordinates. Each point in the figure corresponds to one of the real-world multiplex networks considered in this study. The points from left to right correspond to the following multiplex networks: IPv4/IPv6 Internet, arXiv1-arXiv4, arXiv2-arXiv4, arXiv1-arXiv2, Drosophila1-Drososphila2, Air-Train, C.Elegans2-C.Elegans3, and arXiv1-arXiv5 (Supplemental Material~\cite{SM}, section~I). The $x$-axis shows the average degree exponent $\gamma$ across the two layers of each multiplex. Results in each case are obtained by taking the average value of $D_S$ over $100$ randomized counterparts. Error bars correspond to one standard deviation away from the average. The Pearson correlation coefficient between $D_S$ and $\gamma$ is $r_{D_S, \gamma}=-0.87$ when degree correlation is broken and $r_{D_S, \gamma}=0.48$ when similarity correlation is broken. The dashed lines represent least squares regression lines.}
\label{fig45}
\end{figure}

On the basis of our results, we hypothesize that both degree and similarity correlations matter for the emergence of strong $\mathbf{k}$-core structures. In particular, when the degree distributions of the layers are less heterogeneous, like for the arXiv multiplex network, similarity correlations play a crucial role. On the other hand, when degree distributions are strongly heterogeneous, like in the case of the Internet multiplex network, degree correlations play a crucial role, and the effect of similarities is strongly attenuated (see Supplemental Material~\cite{SM}, section~IV for results from other multiplex network data). This observation is also supported by Figure~\ref{fig45}, which quantifies the difference $D_S$ between the curves of the original and randomized networks of Figures~\ref{fig4}b,f and~\ref{fig5}b,f. The figure also shows $D_S$ for other multiplex systems (considered in Supplemental Material~\cite{SM}, section~IV). We see in Figure~\ref{fig45} that when degree correlation is broken the difference $D_S$ increases as the degree exponent $\gamma$ decreases. On the other hand, when similarity correlation is broken $D_S$ tends to increase with $\gamma$. Figure~\ref{fig45} shows results from different systems that have different parameters (different layer sizes, average degrees, etc.). Therefore, the fact that $D_S$ in Figure~\ref{fig45} is not strictly increasing or decreasing is expected.

\begin{figure}[]
\captionsetup{farskip=0pt}
\includegraphics[width=1\columnwidth]{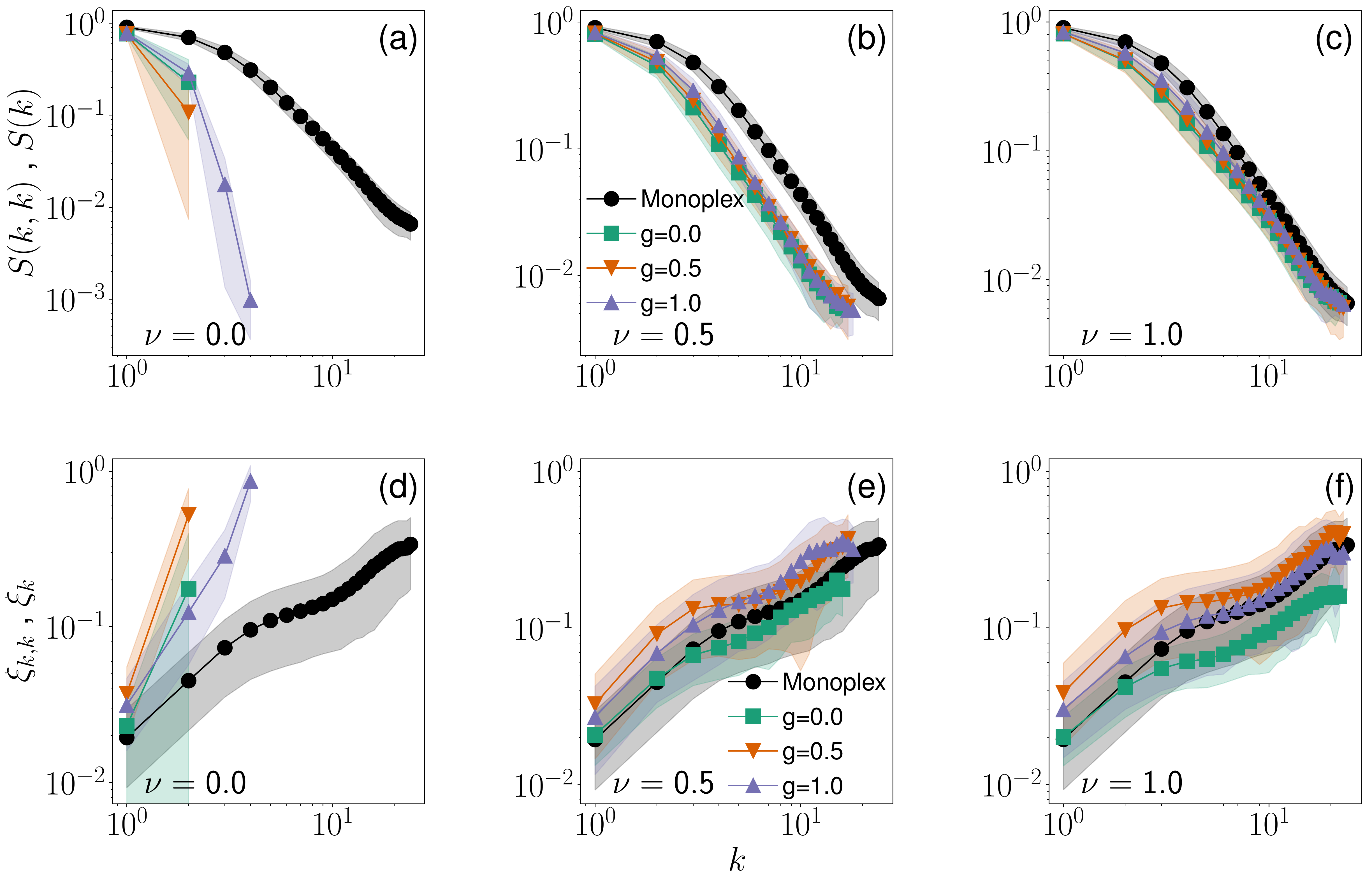}
\caption{ {\bf $\mathbf{k}$-core structure of synthetic multiplex networks.} We study here the effect of degree and angular correlations on the size of the $(k, k)$-core $S(k, k)$ and its coherence $\xi_{k, k}$, in two-layer synthetic multiplex networks constructed according to the Geometric Multiplex Model (Methods Section~\ref{methods_GMM}). Inter-layer degree correlation can be tuned using the parameter $\nu \in [0,1]$, with $\nu = 0$ corresponding to the uncorrelated case, and $\nu = 1$ to the case where
  degrees are maximally correlated. Inter-layer correlation among angular coordinates of nodes is tuned using the model parameter $g \in [0,1]$. When generating network instances according to the GMM, we imposed that
 each layer of the multiplex has $N=10000$ nodes, power-law degree distribution with exponent $\gamma=2.2$, average degree $\bar{k} \approx 6$, and temperature $T=0.5$ (i.e., average clustering coefficient $\bar{c}=0.45$). We consider various combinations of the model parameters $\nu$ and $g$. Results in each case are obtained by taking the average value over $100$ realizations. Shaded areas denote regions corresponding to one standard deviation away from the average.
 ({\bf a})~Relative size $S(k, k)$ of the $(k, k)$-core as a function of the threshold $k$. The curve corresponding to the monoplex is obtained by measuring $S(k)$ for the $k$-core of the individual layers, and then taking the average value.
 ({\bf b} and {\bf c}) Same as in panel a, but for different choices of the model parameters. ({\bf d}, {\bf e} and {\bf f}) We consider the same data as in panels a, b, and c, respectively, but we monitor the metrics of angular coherence $\xi_{k,k}$ and $\xi_k$ as functions of the threshold value $k$.}
\label{fig6}
\end{figure}

\begin{figure}[]
\captionsetup{farskip=0pt}
\includegraphics[width=1\columnwidth]{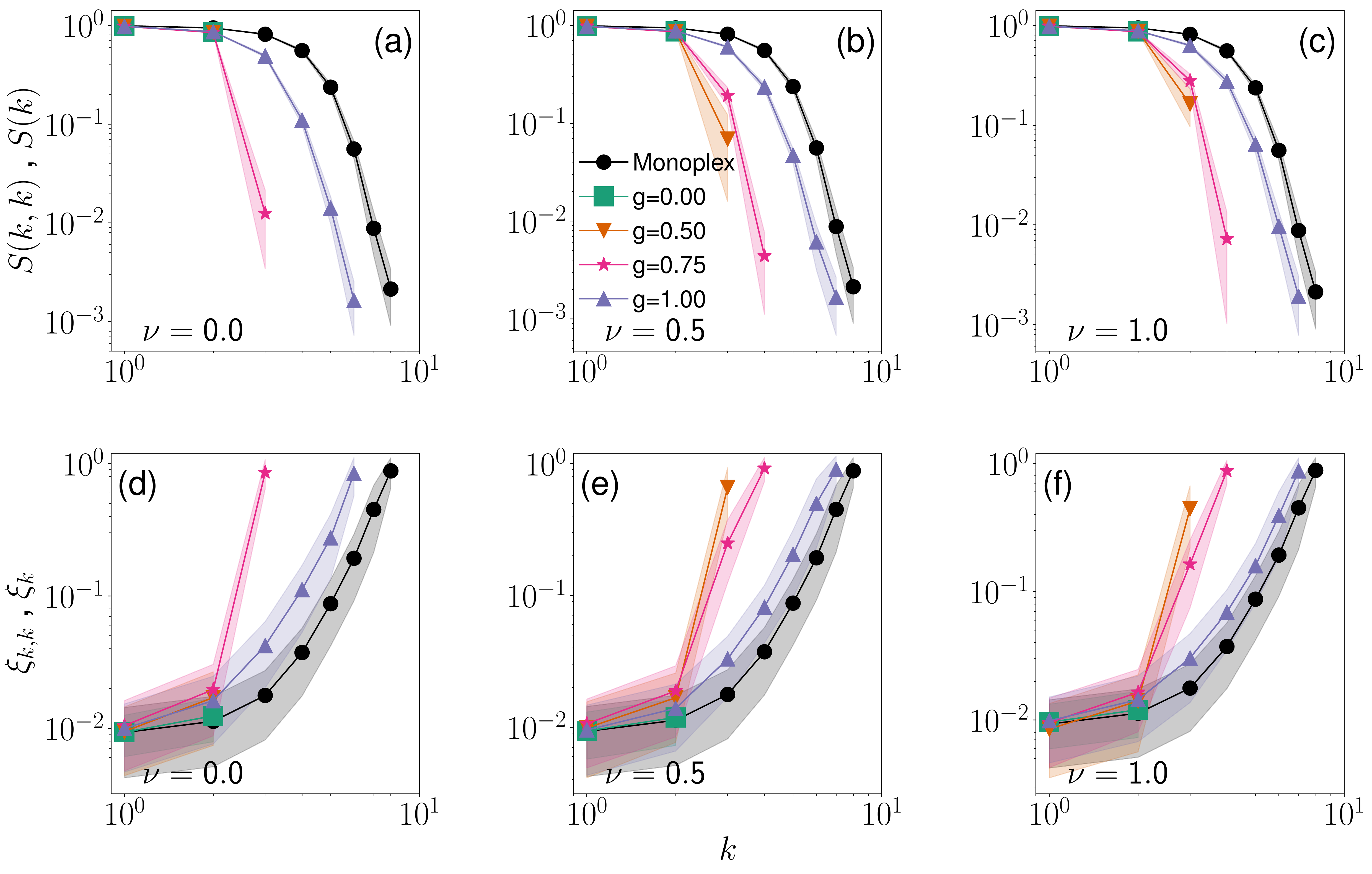}
\caption{{\bf $\mathbf{k}$-core structure of synthetic multiplex networks.} Same as in Figure~\ref{fig6}, but for a different value of the degree exponent $\gamma = 3.5$. Results for model parameter $g=0.75$ are also shown in this figure. All other model parameters are identical to those used in Figure~\ref{fig6}.
}
\label{fig7}
\end{figure}

\begin{figure}[]
\captionsetup{farskip=0pt}
\includegraphics[width=0.78\columnwidth]{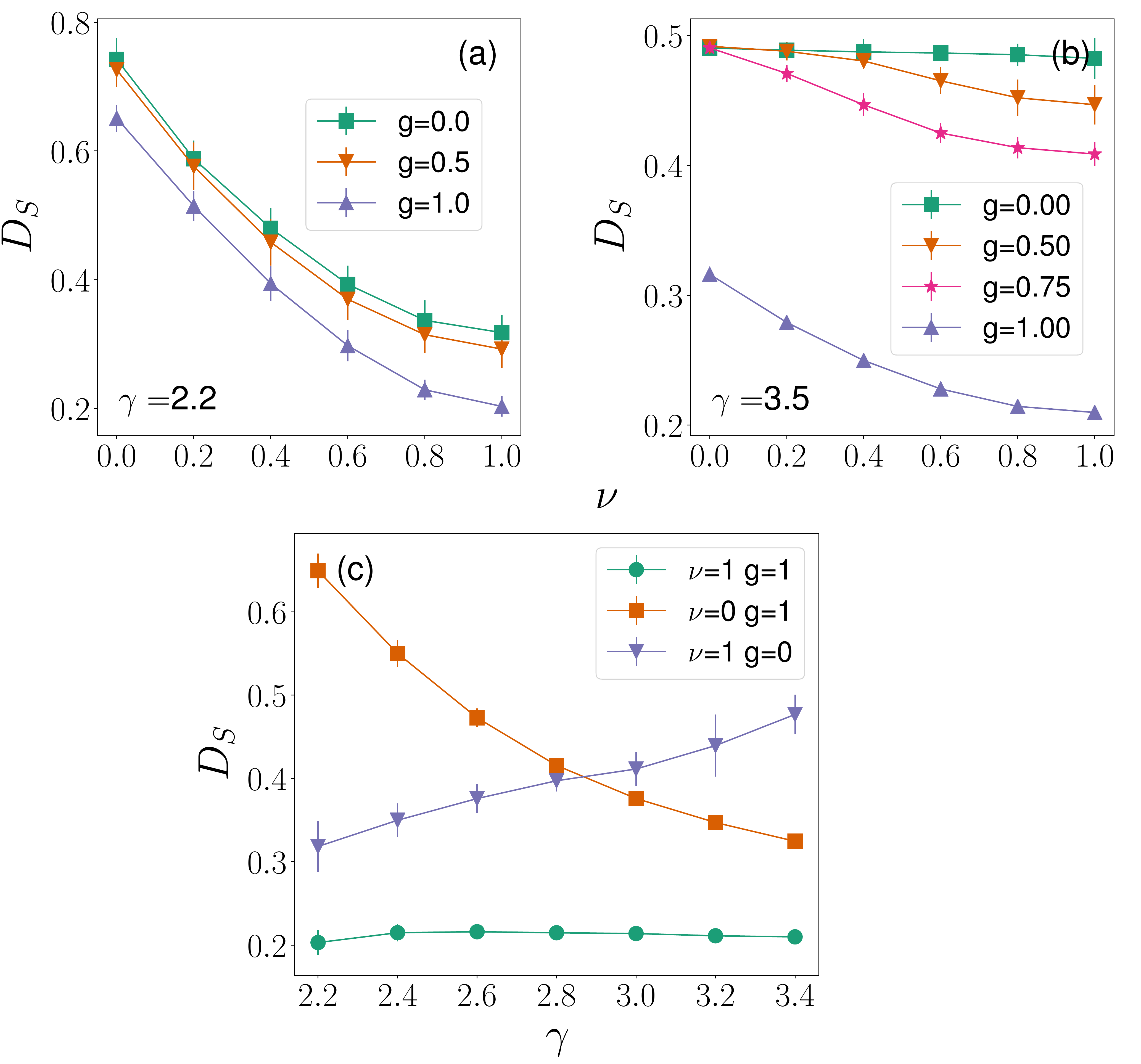}
\caption{{\bf Quantifying the effect of inter-layer degree and similarity correlations in the $\mathbf{k}$-core structure of synthetic multiplex networks.}  ({\bf a} and {\bf b}) Relative difference $D_S=\left[\sum_k S(k)-\sum_k S(k, k)\right]/\sum_k S(k)$ between the monoplex and multiplex relative sizes, $S(k)$ and $S(k, k)$, in two-layer synthetic multiplexes constructed as in Figures~\ref{fig6} and \ref{fig7}. We consider various combinations of the model parameters $\nu$ and $g$.  Results in each case are obtained by taking the average value over $100$ realizations. Error bars correspond to one standard deviation away from the average. Supplemental Material~\cite{SM}, Figure~30 shows also the relative difference $D_\xi=\left[\sum_k \xi_k-\sum_k\xi_{k, k}\right]/\sum_k \xi_k$ between the angular coherences $\xi_k$ and $\xi_{k,k}$ of the networks of panels a and b. Panel ({\bf c}) is the same as panels a and b, but for different values of the degree exponent $\gamma$ and parameters $\nu$ and $g$ as shown in the legend.
}
\label{fig9}
\end{figure}

To test our hypotheses, we rely on synthetic multiplex networks built according to the Geometric Multiplex Model (GMM)~\cite{KleinebergNatPhys}. This model allows to generate single-layer topologies using the $\mathbb{S}^{1}$ model, and control for inter-layer correlation between node degrees and angular coordinates (see Methods section~\ref{methods_GMM}). In Figures~\ref{fig6} and~\ref{fig7}, we study the behavior of the $\mathbf{k}$-core in two-layer synthetic multiplex networks constructed according to the model for different choices of the model parameters (more results can be found in Supplemental Material~\cite{SM}, section~V). We confirm the validity of our claims. Both types of correlations are important for the characterization of the $\mathbf{k}$-core of a multiplex network. Inter-layer degree correlations (measured with $\nu$) are more important than correlations between angular coordinates (measured with $g$) when the degrees of the nodes are broadly distributed. In this case the role of pairwise similarities is much attenuated (see the difference between curves with different $\nu$ versus different $g$ in Figure~\ref{fig6}). If instead,  the network layers are characterized by homogeneous degree distributions, similarity correlations are more important than degree correlations whose role is attenuated (Figure~\ref{fig7}). This effect is also illustrated in Figures~\ref{fig9}a and~\ref{fig9}b, which quantify the differences between the curves of the monoplex and multiplex networks of Figures~\ref{fig6} and~\ref{fig7}, as well as in Figure~\ref{fig9}c, which illustrates a qualitatively similar behavior as the one observed for real networks in Figure~\ref{fig45}. 

The above findings agree with intuition. When the degree distribution of a layer is more heterogeneous there is stronger correlation between higher $k$-shell index values and node degrees (Supplemental Material~\cite{SM}, Figure~2). In other words, the position of similarity of nodes matters less. Thus, inter-layer degree correlations are more important for having a wide $\mathbf{k}$-core structure when the degree distributions of the layers are more heterogeneous. On the other hand, the less heterogeneous is the degree distribution the weaker is the correlation between higher $k$-shell index values and node degrees (Supplemental Material~\cite{SM}, Figure~2). In this case, the position of nodes in the similarity space matters more. Indeed, we have seen that nodes in inner cores have high angular coherence (cf. Figure~\ref{fig1}h). Therefore, inter-layer similarity correlations become more important for having a strong $\mathbf{k}$-core structure  when the degree distributions of the layers are less heterogeneous.

\section*{Discussion and Conclusion}

\begin{figure}[]
\captionsetup{farskip=0pt}
\includegraphics[width=0.8\columnwidth]{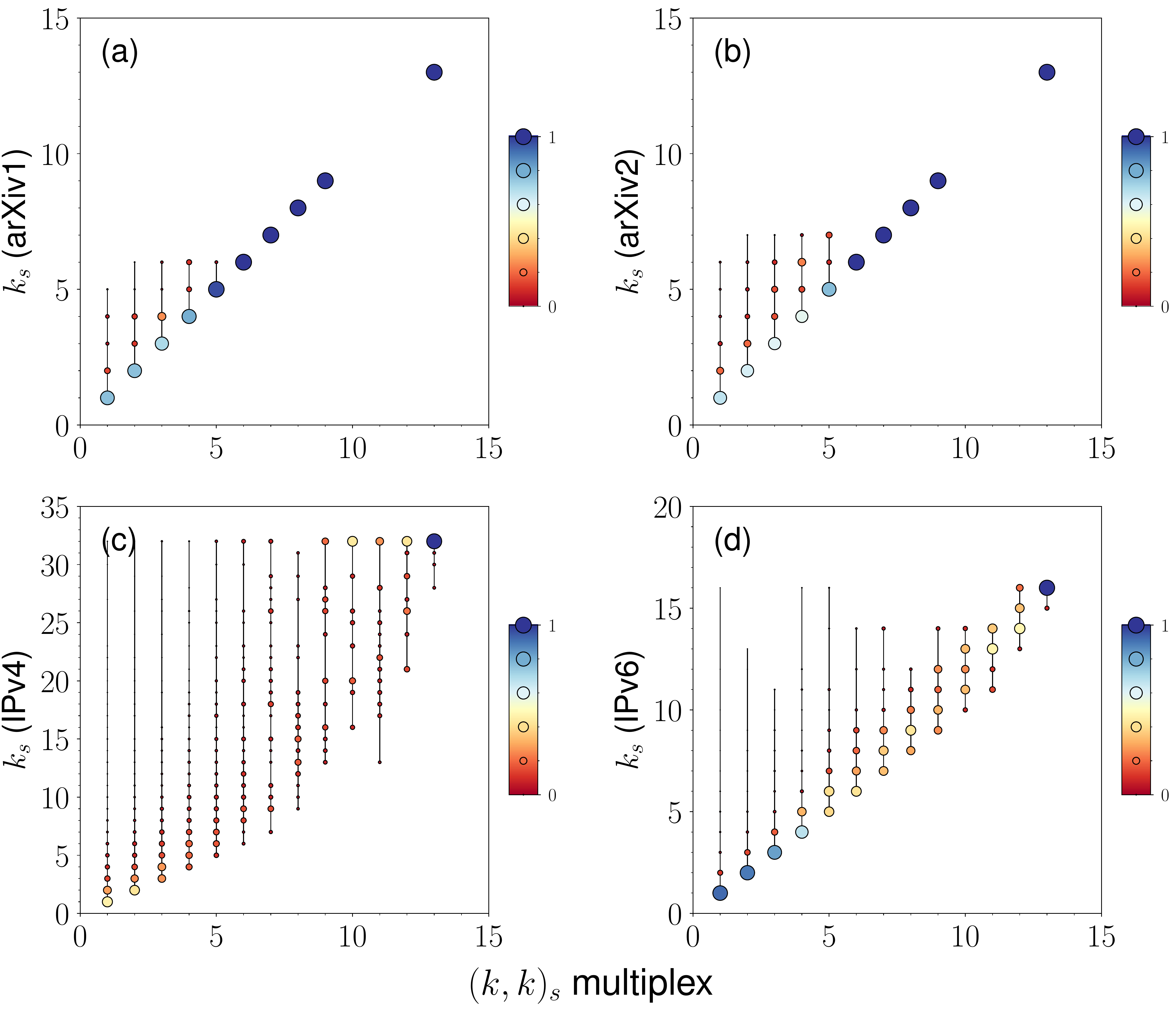}
\caption{{\bf Coreness in single-layers vs. coreness in the multiplex.}  ({\bf a}) Coreness $k_s$ in arXiv1 of the nodes that have coreness $(k, k)_s$ in the multiplex network consisting of arXiv1 and arXiv2. The size of the symbols is proportional to the percentage of nodes with coreness $(k, k)_s$ in the multiplex that have coreness $k_s$ in arXiv1. ({\bf b}) Same as panel a, but for arXiv2. ({\bf c} and {\bf d}) Same as panels a and b, but for the IPv4/IPv6 Internet. 
}
\label{fig11}
\end{figure}

Understanding the principles behind the organization of real-world networks into cores or shells of nodes with increasingly high degree is crucial for better understanding and predicting their structural and dynamical properties, their robustness, and the performance of spreading processes running on top of them. Yet, while the core organization of single-layer networks has been extensively studied in the past, little is known about the core organization of real multiplex networks. In this paper, we performed a systematic characterization of the $\mathbf{k}$-core structure of real-world multiplex networks, and shown that real multiplex networks possess a strong $\mathbf{k}$-core structure that is due to inter-layer correlations. Specifically, we showed that  both degree and similarity correlations between nodes across layers are responsible for the observed strong $\mathbf{k}$-core structures. The more heterogeneous are the degree distributions of the layers, the more pivotal is the role of degree correlations. On the other hand, the more homogeneous are the degree distributions, the more crucial is the role of similarity correlations. We reached our conclusions by taking advantage of network hyperbolic embedding, and showed that such a geometric description of networks provides a simple framework to naturally understand and characterize the $\mathbf{k}$-core structure of real-world multiplex networks. As the core organization of a network is intimately related to the behavior of spreading phenomena~\cite{Kitsak2010}, our results open the door for a geometric perspective in understanding and predicting the efficiency of spreading processes and the location of influential spreaders in real multiplex networks. Indeed, the wide $\mathbf{k}$-core structure found in real multiplex systems, explained by inter-layer geometric correlations, suggests that there are nodes, located into inner $\mathbf{k}$-cores, which could potentially act as efficient spreaders in all layers of the multiplex simultaneously. For instance, we see in Figure~\ref{fig11} that in the Internet and arXiv multiplexes nodes with high $(k, k)$-shell index in the multiplex have also high $k$-shell index in the individual layers. Further, in contrast to arXiv, where the nodes in the most inner $k$-shells of the individual layers belong also to the most inner $(k, k)$-shells of the multiplex, in the IPv4/IPv6 Internet there are nodes with high $k$-shell index values in the individual layers but not in the multiplex. This suggests that there are also nodes that could potentially be efficient spreaders in the individual layers but not in the multiplex. We leave such investigations for future work.

\section*{Acknowledgments}
 F.R. acknowledges support from the National Science Foundation (CMMI-1552487) and the U.S. Army Research Office (W911NF-16-1-0104).

\section*{Methods}

\subsection{\bf Cores and shells}
\label{methods_core}

The $k$-core of a single-layer network is the maximal subgraph of the network in which all vertices have degree at least $k$. The $k$-core is identified by iteratively removing all nodes with degree less than $k$, recalculating the degrees of all the remaining nodes, and continuing with the iterative scheme till there are no nodes with degree less than $k$. By definition, all nodes in the $(k+n)$-core, with $n \geq 0$, are necessarily part of the $k$-core.  The nodes that belong to the $k$-core but not to the $(k+1)$-core form the $k$-shell of the network, and they are said to have $k$-shell index, or coreness, $k_\textnormal{s}=k$.  The relative size $S(k)$ of the $k$-core is
\begin{equation}
\label{eq:s_k1}
S(k) = \frac{N_k}{N},
\end{equation}
where $N_{k}$ is the number of nodes that belong to the $k$-core, and $N$ is the total number of nodes in the network.

In a multiplex system of $L$~layers, the $\mathbf{k}$-core, with $\mathbf{k}  = (k_1, \ldots ,  k_\ell , \ldots, k_L)$, is the set of the subgraphs, one for each layer, remaining after the following pruning procedure is performed~\cite{AzimiKcore2014PRE}: all nodes whose degree in at least one layer $\ell$ is less than $k_\ell$ are removed from the system; the degree of all nodes in all layers is recomputed; the pruning continues iteratively until no node remains such that its degree in layer $\ell$ is less than the threshold $k_\ell$.  By definition, the subgraphs belonging to the $\mathbf{k}$-core share the same set of nodes. Further, the $(\mathbf{k} + \mathbf{n})$-core of a multiplex, with $\mathbf{n} = (n_1, \ldots, n_\ell, \ldots, n_L)$ where $n_\ell \geq 0$ for all $\ell =1, \ldots, L$, is necessarily a subset of the $\mathbf{k}$-core of the multiplex. Similar to single-layer networks one can also define $\mathbf{k}$-shells. Figure~\ref{fig3}c in the main text illustrates the $(k, k)$-shells in the considered arXiv multiplex, i.e., the sets of nodes that belong to the $(k, k)$-core but not to the $(k+1, k+1)$-core of the system, $k=1, 2, \dots, 13$.

The relative size $S(\mathbf{k})$ of the $\mathbf{k}$-core is
\begin{equation}
\label{eq:s_k2}
S(\mathbf{k}) = \frac{N_\mathbf{k}}{N},
\end{equation}
where $N_\mathbf{k}$ is the number of nodes belonging to the $\mathbf{k}$-core, and $N$ is the number of common nodes between the layers of the multiplex.

\subsection{Network randomization}
\label{methods_rand}

\subsubsection{Single-layer randomization}
In Figure~\ref{fig1} of the main text, we employed a degree-preserving clustering-decreasing randomization procedure that works as follows. We select a random pair of links $(i,j)$ and $(s,t)$ in the network, and rewire them to $(i,t)$ and $(s,j)$, provided that none of these links already exist in the network and that the rewiring decreases the average clustering coefficient $\bar{c}$~\cite{Dorogovtsev10-book} in the network. If these two conditions are met, then the rewiring is accepted, otherwise it is not accepted, and a new pair of links is selected. This way each accepted rewiring step preserves the degree distribution in the network, and decreases its average clustering. We repeat the rewiring steps till we reach desired pre-defined values of the average clustering coefficient $\bar{c}$, as shown in the legends of Figures~\ref{fig1}b and~\ref{fig1}f.

\subsubsection{Multiplex randomization}
 In Figure~\ref{fig3} of the main text, we employed a node label reshuffling procedure that destroys all correlations between two layers of a multiplex. Specifically, we randomly reshuffled the labels of the nodes of one layer, i.e., we interchanged the label of each node in that layer with the label of a randomly selected node from the same layer. This process randomly reshuffles the trans-layer node-to-node mappings without altering the layer topology. 

\subsection{\bf $\mathbb{S}^{1}$ model}
\label{methods_S1}

Each node $i$ in the $\mathbb{S}^{1}$ model has hidden variables $\kappa_i, \theta_i$. The hidden variable $\kappa_i$ is the node's expected degree in the resulting network, while $\theta_i$ is the angular (similarity) coordinate of the node on a circle of radius $R=N/(2\pi)$, where $N$ is the total number of nodes. To construct a network with the $\mathbb{S}^{1}$ model that has size $N$, average node degree $\bar{k}$, power law degree distribution with exponent $\gamma > 2$, and temperature $T \in [0,1)$, we perform the following steps:
\begin{enumerate}
\item[i.] Sample the angular coordinates of nodes $\theta_i$, $i=1,2,\ldots,N$, uniformly at random from $[0, 2\pi]$, and their hidden variables $\kappa_{i}$, $i=1,2,\ldots,N$, from the probability density function
\begin{equation}
\rho(\kappa) = (\gamma-1) \kappa_0^{\gamma-1} \kappa^{-\gamma},
\end{equation}
where $\kappa_0=\bar{k}(\gamma-2)/(\gamma-1)$ is the expected minimum node degree;
\item[ii.] Connect every pair of nodes $i,j$ with probability
\begin{equation}
\label{p_s1}
p(\chi_{ij})=\frac{1}{1+\chi_{ij}^{1/T}},
\end{equation}
where $\chi_{ij}= R \Delta \theta_{ij}/(\mu\kappa_i\kappa_j)$ is the effective distance between $i$ and $j$, $\Delta \theta_{ij}=\pi - | \pi -|\theta_i - \theta_j||$ is the angular distance, and $\mu =\sin{T \pi}/(2\bar{k}T\pi)$ is derived from the condition that the expected degree in the network is indeed $\bar{k}$.
\end{enumerate}
The $\mathbb{S}^{1}$ model is isomorphic to hyperbolic geometric graphs ($\mathbb{H}^{2}$ model) after transforming the expected node degrees $\kappa_i$ to radial coordinates $r_i$ via
\begin{equation}
\label{kappa_i}
r_i = R_\textnormal{H} - 2 \ln \frac{\kappa_i}{\kappa_0},
\end{equation} 
where $R_\textnormal{H}$ is the radius of the hyperbolic disc where all nodes reside,
\begin{equation}
\label{R}
R_{\textnormal{H}}=2\ln{\frac{N}{c}},
\end{equation}
while $c=\bar{k}\frac{\sin{T\pi}}{2T}\left(\frac{\gamma-2}{\gamma-1}\right)^2$. After this change of variables the connection probability in~(\ref{p_s1}) becomes
\begin{equation}
\label{p_h2}
p(x_{ij})=\frac{1}{1+e^{\frac{1}{2T}(x_{ij}-R_\textnormal{H})}}, 
\end{equation}
where $x_{ij} = r_i+r_j+2\ln{(\Delta \theta_{ij}/2)}$ is approximately the hyperbolic distance between nodes $i, j$~\cite{Krioukov2010PRE}. 

\subsection{Hyperbolic embedding}
\label{methods_embedding}

The hyperbolic embeddings of all considered real-world networks have been obtained in~\cite{KleinebergNatPhys} using the HyperMap embedding method~\cite{hypermap_cn}. The method is based on maximum likelihood estimation. On its input it takes the network adjacency matrix  $A$. The generic element of the matrix is $A_{ij}=A_{ji}=1$ if there is a link between nodes $i$ and $j$, and $A_{ij}=A_{ji}=0$ otherwise. The embedding infers radial and angular coordinates, respectively indicated as  $r_i$ and $\theta_i$, for all nodes $i \leq N$. The radial coordinate $r_i$ is related to the observed node degree $k_i$ as
\begin{equation}
r_i \sim \ln{N}-2\ln{k_i} \; .
\end{equation}
The angular coordinates of nodes are found by maximizing the likelihood
\begin{equation}
\label{eq:likelihood}
\mathcal L=\prod_{1 \leq j < i \leq N} p(x_{ij})^{A_{ij}}\left[1-p(x_{ij})\right]^{1-A_{ij}}.
\end{equation}
The product in the above relation goes over all node pairs $i, j$ in the network, $x_{ij}$ is the hyperbolic distance between pair $i, j$~\cite{Krioukov2010PRE} and $p(x_{ij})$ is the connection probability in Eq.~(\ref{p_h2}).

\subsection{Angular coherence}
\label{methods_coherence}

\subsubsection{Single-layer networks}

To quantify how similar are the angular coordinates of nodes in the $k$-cores, we use angular coherence, a metric previously used to quantify the extent to which nodes within the same community have similar angular coordinates~\cite{FaqeehPRL}. We define the angular coherence of a $k$-core as the module $0 \leq \xi_k \leq 1$, given by
\begin{equation}
\label{eq:coherence}
\xi_k e^{i \phi_k} = \frac{1}{N_k} \sum_{j \in k\textrm{-core}} e^{i \theta_j},
\end{equation}
where the sum is taken over the set of nodes that belong to the $k$-core, $N_{k}$ is the number of nodes that belong to the $k$-core, and $\theta_j$ is the angular coordinate of node $j$. The angular coherence resembles the order parameter of the Kuramoto model that captures the coherence of oscillators~\cite{Kuramoto}. The higher is the $\xi_k \in [0,1]$ the more localized in the similarity space are the nodes of the $k$-core. At $\xi_k=1$ all nodes have the same angular coordinates, while at $\xi_k=0$ nodes are uniformly distributed in $[0, 2\pi]$. $\phi_k$ in Eq.~(\ref{eq:coherence}) can be seen as the $k$-core's ``angular coordinate", i.e., it is a measure of where the $k$-core is mostly concentrated along the angular similarity direction. We note that the angular coherence of a $k$-core is an average metric, taken over the nodes that belong to the $k$-core. Therefore, the value of $\xi_k$ does not depend on the number of nodes $N_k$ that belong to the $k$-core.

\subsubsection{Multiplex networks}
For two-layer multiplex networks, we define the angular coherence of the nodes belonging to the $(k, k)$-core as the module $0 \leq \xi_{k, k} \leq 1$, given by averaging the angular coherences of the corresponding nodes in the individual layers,
\begin{equation}
\label{eq:coherence_m}
\xi_{k, k} e^{i \phi_{k, k}} = \frac{1}{2} \sum_{\ell=1}^{2}  
\Bigg(
\frac{1}{N_{k, k}} 
\sum_{j \in (k, k)\textrm{-core}} e^{i \theta^{\ell}_j}
\Bigg),
\end{equation}
where $N_{k, k}$ is the number of nodes belonging to the $(k, k)$-core, and $\theta^{\ell}_j$ is the angular coordinate of node $j$ in layer $\ell=1,2$. Similar to $\xi_k$, $\xi_{k,k}$ does not depend on the number of nodes $N_{k, k}$ that belong to the $(k, k)$-core.

\subsection{\bf Inter-layer similarity}
\label{methods_similarity}

\subsubsection{Degree correlation}
Degree correlation between two layers of a multiplex network is quantified using the Pearson correlation coefficient~\cite{KleinebergNatPhys}
\begin{equation}
\label{r}
r_{k,k'}=\frac{\textnormal{cov}(k, k')}{\sigma_k \sigma_{k'}},
\end{equation}
where $\textnormal{cov}(X, X')$ denotes the covariance between two random variables $X$ and $X'$ and $\sigma_x$ denotes the standard deviation of random variable $X$.  $r_{k, k'}$ takes values in $[-1, 1]$ and is computed across the nodes that are common in the two layers. For $r_{k, k'}=1$ the degrees of the nodes in the two layers are fully correlated, for $r_{k,k'}=0$ they are uncorrelated, while for $r_{k, k'}=-1$ they are fully anti-correlated.

\subsubsection{Angular correlation}
Angular correlation between the two layers of a multiplex is quantified using the normalized mutual information~\cite{KleinebergNatPhys}
\begin{equation}
\label{NMI}
\textnormal{NMI}_{\theta,\theta'}=\frac{\textnormal{MI}(\theta ; \theta')}{\textnormal{max}\{\textnormal{MI}(\theta ; \theta), \textnormal{MI}(\theta' ; \theta') \}},
\end{equation}
where \textnormal{MI} is the mutual information, computed using the method proposed in Ref.~\cite{kraskov2004estimating}. $\textnormal{NMI}_{X, X'}$ takes values in $[0, 1]$ and is computed across the common nodes in the two layers. $\textnormal{NMI}_{X, X'}=0$ means no correlation between $X$ and $X'$, while $\textnormal{NMI}_{X, X'}=1$ means perfect correlation.

\subsubsection{Edge overlap}
The edge overlap $\mathcal{O}$ between two layers is given by 
\begin{equation}
\label{o}
\mathcal{O}=\frac{\sum_{i>j} A_{ij} A'_{ij}}{\textnormal{min} \{ \sum_{i>j} A_{ij} , \sum_{i>j} A'_{ij} \}},
\end{equation}
where $A$ and $A'$ are the adjacency matrices of the two layers. The numerator in (\ref{o}) is the number of overlapping links between the two layers, while the denominator is the maximum possible number of overlapping links. 

\subsection{\bf Geometric Multiplex Model}\label{methods_GMM}

The Geometric Multiplex Model (GMM) generates single-layer topologies using the $\mathbb{S}^{1}$ model (Methods section~\ref{methods_S1}), and allows for degree and angular coordinate correlations across the layers. Specifically, correlations can be tuned by varying the model parameters $\nu \in [0,1]$ (degree correlations) and $g \in [0,1]$ (angular correlations)~\cite{KleinebergNatPhys}. Degree (angular) correlations are maximized at $\nu \to 1$ ($g \to 1$), while at $\nu \to 0$ ($g \to 0$) there are no degree (angular) correlations. The GMM implementation is available at~\cite{GMMCODE}.

\bibliography{refs}

\newpage

\onecolumngrid

\end{document}